\documentclass[12pt]{article}
\usepackage{graphicx}

      \usepackage[cp1251]{inputenc} 
      \usepackage[russian]{babel}   

\begin{document}
 \noindent {\footnotesize\it Astronomy Letters, 2014, Vol. 40, No. 7, pp. 389--397.}
 \newcommand{\dif}{\textrm{d}}

 \noindent
 \begin{tabular}{llllllllllllllllllllllllllllllllllllllllllllll}
 & & & & & & & & & & & & & & & & & & & & & & & & & & & & & & & & & & & & & \\\hline\hline
 \end{tabular}

 \vskip 1.0cm

 \centerline{\bf Determination of Galactic Rotation Parameters and the}
 \centerline{\bf Solar Galactocentric Distance $R_0$ from 73 Masers}
 \bigskip
 \centerline{\bf V.V. Bobylev$^{1,2}$ and A.T. Bajkova$^1$}
 \bigskip
 \centerline{\small \it $^1$Pulkovo Astronomical Observatory, St. Petersburg,  Russia}
 \centerline{\small \it $^2$Sobolev Astronomical Institute, St. Petersburg State University, Russia}
 \bigskip
 \bigskip
{\bf Abstract}—We have determined the Galactic rotation parameters
and the solar Galactocentric distance $R_0$ by simultaneously
solving Bottlinger’s kinematic equations using data on masers with
known line-of-sight velocities and highly accurate trigonometric
parallaxes and proper motions measured by VLBI. Our sample
includes 73 masers spanning the range of Galactocentric distances
from 3 to 14 kpc. The solutions found are
 $\Omega_0 = 28.86\pm0.45$~km s$^{-1}$ kpc$^{-1}$ ,
 $\Omega'_0 = -3.96\pm0.09$~km s$^{-1}$ kpc$^{-2}$,
 $\Omega''_0 =0.790\pm0.027$~km s$^{-1}$ kpc$^{-3},$ and
 $R_0=8.3\pm0.2$ kpc.
In this case, the linear rotation velocity at the solar distance
$R_0$ is $V_0=241\pm7$~km s$^{-1}$. Note that we have obtained the
$R_0$ estimate, which is of greatest interest, from masers for the
first time; it is in good agreement with the most recent estimates
and even surpasses them in accuracy.


\section*{INTRODUCTION}
Both kinematic and geometric characteristics are important for
studying the Galaxy, with the solar Galactocentric distance $R_0$
being the most important among them. Various data are used to
determine the Galactic rotation parameters. These include the line
of sight velocities of neutral and ionized hydrogen clouds with
their distances estimated by the tangential point method (Clemens
1985; McClure-Griffiths and Dickey 2007; Levine et al. 2008),
Cepheids with the distance scale based on the period–luminosity
relation, open star clusters and OB associations with photometric
distances (Mishurov and Zenina 1999; Rastorguev et al. 1999;
Zabolotskikh et al. 2002; Bobylev et al. 2008; Mel’nik and Dambis
2009), and masers with their trigonometric parallaxes measured by
VLBI (Reid et al. 2009a; McMillan and Binney 2010; Bobylev and
Bajkova 2010; Bajkova and Bobylev 2012).

The solar Galactocentric distance $R_0$ is often assumed to be
known in a kinematic analysis of data, because not all of the
kinematic data allow $R_0$ to be reliably estimated. In turn,
different (including direct) methods of analysis give different
values of $R_0$.

Reid (1993) published a review of the $R_0$ measurements made by
then by various methods. He divided all measurements into primary,
secondary, and indirect ones and obtained the ``best value'' as a
weighted mean of the published measurements over a period of 20
years: $R_0=8.0\pm0.5$~kpc. Nikiforov (2004) proposed a more
complete three-dimensional classification in which the type of
$R_0$ determination method, the method of finding the reference
distances, and the type of reference objects are taken into
account. Taking into account the main types of errors and
correlations associated with the classes of measurements, he
obtained the ``best value'' $R_0=7.9\pm0.2$~kpc by analyzing the
results of various authors published between 1974 and 2003.

Based on 52 results published between 1992 and 2010, Foster and
Cooper (2010) obtained the mean $R_0=8.0\pm0.4$~kpc. The paper by
Malkin (2013) is devoted to studying the dependence of these data
on the date of publication. He argues for the absence of a
statistically significant ``bandwagon'' effect and found the mean
value to be close to $R_0=8.0$~kpc. Francis and Anderson (2013)
gave a summary of 135 publications devoted to the $R_0$
determination between 1918 and 2013. They concluded that the
results obtained after 2000 give a mean value of $R_0$ close to
8.0~kpc.

We have some experience of determining $R_0$ by simultaneously
solving Bottlinger’s kinematic equations with the Galactic
rotation parameters. To this end, we used data on open star
clusters (Bobylev et al. 2007) distributed within about 4 kpc of
the Sun. Clearly, using masers belonging to regions of active star
formation and distributed in a much wider region of the Galaxy for
this purpose is of great interest. However, the first such
analysis for a sample of 18 masers performed by McMillan and
Binney (2010) showed the probable value of $R_0$ to be within a
fairly wide range, 6.7--8.9~kpc. At present, the number of masers
with measured trigonometric parallaxes has increased (about 80),
which must lead to a significant narrowing of this range.

The goal of this paper is to determine the Galactic rotation
parameters and the distance $R_0$ using data on masers with
measured trigonometric parallaxes.

\section*{METHOD}
Here, we use a rectangular Galactic coordinate system with the
axes directed away from the observer toward the Galactic center
$(l$=$0^\circ$, $b$=$0^\circ,$ the $X$ axis), in the direction of
Galactic rotation ($(l$=$90^\circ$, $b$=$0^\circ,$ the $Y$ axis),
and toward the north Galactic pole ($b=90^\circ,$ the $Z$ axis).

The method of determining the kinematic parameters consists in
minimizing a quadratic functional $F:$
  \begin{equation}
   \begin{array}{rll}
 \min~F=
 \sum_{j=1}^N [w_r^j (V_r^j-\hat{V}_{r}^j)]^2+
 \sum_{j=1}^N [w_l^j (V_l^j-\hat{V}_{l}^j)]^2+
 \sum_{j=1}^N [w_b^j (V_b^j-\hat{V}_{b}^j)]^2
  \label{Functional}
 \end{array}
\end{equation}
provided that the following constraints derived from Bottlinger's
formulas with an expansion of the angular velocity of Galactic
rotation $\Omega$ into a series to terms of the second order of
smallness with respect to $r/R_0$ are fulfilled:
 \begin{equation}
 \begin{array}{lll}
 V_r=-u_\odot\cos b\cos l-v_\odot\cos b\sin l-w_\odot\sin b\\
 +R_0(R-R_0)\sin l\cos b \Omega'_0+0.5R_0 (R-R_0)^2 \sin l\cos b \Omega''_0+r K \cos^2 b,
\label{EQ-1}
 \end{array}
 \end{equation}
 \begin{equation}
 \begin{array}{lll}
 V_l= u_\odot\sin l-v_\odot\cos l+(R-R_0)(R_0\cos l-r\cos b) \Omega'_0\\
  +(R-R_0)^2 (R_0\cos l - r\cos b)0.5\Omega''_0 - r \Omega_0 \cos b,
 \label{EQ-2}
 \end{array}
 \end{equation}
 \begin{equation}
 \begin{array}{lll}
 V_b=u_\odot\cos l \sin b+v_\odot\sin l \sin b-w_\odot\cos b\\
    -R_0(R-R_0)\sin l\sin b\Omega'_0-0.5R_0(R-R_0)^2\sin l\sin b\Omega''_0-r K \cos b\sin b,
\label{EQ-3}
 \end{array}
 \end{equation}
where $N$ is the number of objects used; $j$ is the current object
number; $V_r$ and $V_l,$ $V_b$ are the model values of the
three-dimensional velocity field: the line-of-sight velocity and
the proper motion velocity components in the $l$ and $b$
directions, respectively; $V_l=4.74 r \mu_l\cos b$, $V_b=4.74 r
\mu_b$ are the measured components of the velocity field (data),
with $\hat{V}_{r}^j, \hat{V}_{l}^j$ and $\hat{V}_{b}^j$, where the
coefficient 4.74 is the quotient of the number of kilometers in an
astronomical unit and the number of seconds in a tropical year;
$w_r^j, w_l^j,w_b^j$ are the weight factors; $r$ is the
heliocentric distance of the star calculated via the measured
parallax $\pi,$ $r=1/\pi$; the star's proper motion components
$\mu_l\cos b$ and $\mu_b$ are in mas yr$^{-1}$ (milliarcseconds
per year), the line-of-sight velocity $V_r$ is in km s$^{-1}$;
$u_\odot,v_\odot,w_\odot$ are the stellar group velocity
components relative to the Sun taken with the opposite sign (the
velocity $u$ is directed toward the Galactic center, $v$ is in the
direction of Galactic rotation, $w$ is directed to the north
Galactic pole), when needed we assume $w_\odot$ to be 7~km
s$^{-1}$, because it is poorly determined from distant objects;
$R_0$ is the Galactocentric distance of the Sun; $R$ is the
distance from the star to the Galactic rotation axis,
 \begin{equation}
 \begin{array}{rll}
 R^2=r^2\cos^2 b-2R_0 r\cos b\cos l+R^2_0.
 \label{RR}
 \end{array}
 \end{equation}
$\Omega_0$ is the angular velocity of rotation at the distance
$R_0;$ the parameters $\Omega'_0$ and $\Omega''_0$ are the first
and second derivatives of the angular velocity with respect to
$R,$ respectively; $K$ is the Oort constant that describes the
local expansion/contraction of the stellar system.

The weight factors in functional (1) are assigned according to the
following expressions (for simplification, we omitted the index
$i$):
 \begin{equation}
 \begin{array}{rll}
 w_r=       S_0/\sqrt {S_0^2+\sigma^2_{V_r}},\qquad
 w_l=\beta  S_0/\sqrt {S_0^2+\sigma^2_{V_l}},\qquad
 w_b=\gamma S_0/\sqrt {S_0^2+\sigma^2_{V_b}},
 \label{WESA}
 \end{array}
 \end{equation}
where $S_0$ denotes the dispersion averaged over all observations,
which has the meaning of a ``cosmic'' dispersion taken to be
8~km~s$^{-1}$;  $\beta=\sigma_{V_r}/\sigma_{V_l}$ and
$\gamma=\sigma_{V_r}/\sigma_{V_b}$ are the scale factors, where
$\sigma_{V_r},\sigma_{V_l}$ and $\sigma_{V_b}$ denote the velocity
dispersions along the line of sight, the Galactic longitude, and
the Galactic latitude, respectively. The system of weights (6) is
close to that from Mishurov and Zenina (1999). We take
$\beta=\gamma=1,$ as the initial approximation; below we describe
the procedure of refining these parameters.

The errors of the velocities $V_l$ and $V_b$ are calculated from
the formula
 \begin{equation}
 \sigma_{(V_l,V_b)}=4.74r\sqrt{\mu^2_{l,b}\Biggl({\sigma_r\over r}\Biggr)^2+\sigma^2_{\mu_{l,b}}}.
  \label{Errors}
 \end{equation}
In addition to the system of weights (6) described above, we also
use the case of unit weights where $w_r=w_l=w_b=1$ for comparison.

The problem of optimizing functional (1), given Eqs. (2)--(4), is
solved numerically for the eight unknown parameters  $u_\odot$,
 $v_\odot$, $w_\odot$, $\Omega_0$, $\Omega'_0$, $\Omega''_0$, $K$ and $R_0$
from a necessary condition for the existence of an extremum. A
sufficient condition for the existence of an extremum in a
particular domain is the positive definiteness of the Hessian
matrix composed of the elements $\{a_{i,j}\}=d^2F/dx_i dx_j$,
where $x_i (i=1,...,8)$ denote the sought-for parameters,
everywhere in this domain. We calculated the Hessian matrix in a
wide domain of parameters or, more specifically, $\pm50\%$ of the
nominal values of the parameters.

Our analysis of the Hessian matrix for both cases of weighting
showed its positive definiteness, suggesting the existence of a
global minimum in this domain and, as a consequence, the
uniqueness of the solution. As an example, Fig.~1 shows the
two-dimensional residuals, or the square root of the functional
$F,$ with one of the measurements being specified by the parameter
$R_0$ and one of the parameters $u_\odot, v_\odot, w_\odot,
\Omega_0, \Omega'_0,$ and $\Omega''_0$, acting as the second
measurement, provided that the remaining parameters from the
series are fixed at the level of the solution obtained. The
presented pictures clearly demonstrate a global minimum in a wide
domain of parameters. In the case of unit weight factors, the
Hessian matrix is also positively defined far beyond this domain.
However, as will be shown below, the adopted weighting allowed the
accuracy of the solutions obtained to be increased.

We estimated the errors of the sought-for parameters through Monte
Carlo simulations. The errors were estimated by performing 100
cycles of computations. For this number of cycles, the mean values
of the solutions virtually coincide with the solutions obtained
purely from the initial data, i.e., without adding any measurement
errors.

\section*{DATA}
Based on published data, we gathered information about the
coordinates, line-of-sight velocities, proper motions, and
trigonometric parallaxes of Galactic masers measured by VLBI with
an error, on average, less than 10\%. These masers are associated
with very young objects, protostars of mostly high masses located
in regions of active star formation. The proper motions and
trigonometric parallaxes of the masers are absolute, because they
are determined with respect to extragalactic reference objects
(quasars).

One of the projects to measure the trigonometric parallaxes and
proper motions is the Japanese VERA (VLBI Exploration of Radio
Astrometry) project devoted to the observations of H$_2$O masers
at 22.2 GHz (Hirota et al. 2007) and a number of SiO masers (which
are very few among young objects) at 43 GHz (Kim et al. 2008).

Methanol (CH$_3$OH, 6.7 and 12.2 GHz) and H$_2$O masers are
observed in the USA on VLBA (Reid et al. 2009a). Similar
observations are also being carried out within the framework of
the European VLBI network (Rygl et al. 2010), in which three
Russian antennas are involved: Svetloe, Zelenchukskaya, and
Badary. These two programs enter into the BeSSeL project\footnote
{http://www3.mpifr-bonn.mpg.de/staff/abrunthaler/BeSSeL/index.shtml}
(Bar and Spiral Structure Legacy Survey, Brunthaler et al. 2011).

The VLBI observations of radio stars in continuum at 8.4 GHz are
being carried out with the same goals (Torres et al. 2009; Dzib et
al. 2011). Radio sources located in the local (Orion) arm
associated with young low-mass protostars are observed within the
framework of this program.

Information about 44 masers (coordinates, line-of-sight
velocities, proper motions, and parallaxes) is presented in
Bajkova and Bobylev (2012). Subsequently, a number of new
measurements have been published by various authors. These data on
31 sources are provided in Bobylev and Bajkova (2013b). The
initial data on several new masers published later (Zhang et al.
2013) are given in Table 1.

It should be noted that the measurements for a number of masers
were performed by various authors several times. For example, for
the Onsala 2 region (G75.78$+$0.34), we used the measurements from
Xu et al. (2013). Zhang et al. (2013) point out that the
measurement of the distance to the G048.60$+$0.02 region by
Nagayama et al. (2011) is probably erroneous and, therefore, we
did not consider this measurement. Five sources were rejected
according to the 3$\sigma$ criterion. As a result, our sample for
a kinematic analysis in the range of distances $R$ from 3 to 14
kpc contains 73 masers.

      \begin{table}[t]
      \begin{center}
      \caption{Initial data on the masers}
      {\small
  \label{t0}
      \begin{tabular}{|l|r|r|r|r|r|r|r|r|r|r|r|}      \hline
      Source & $\alpha$ & $\delta$ & $\pi(\sigma_\pi)$ &
      $\mu^*_\alpha (\sigma_{\mu_\alpha})$ & $\mu_\delta(\sigma_{\mu_\delta})$ &
      $V_r(\sigma_{V_r})$ & Ref \\\hline

G43.16+0.01 & $287^\circ.5559$ & $ 9^\circ.1036$ & $.090(.006)$ & $-2.48(.15)$ & $-5.27(.13)$ & $11.0(5.0)$ & (1) \\
G48.60+0.02 & $290^\circ.1299$ & $13^\circ.9237$ & $.093(.005)$ & $-2.89(.13)$ & $-5.50(.13)$ & $18.0(5.0)$ & (1) \\
 \hline
      \end{tabular}}
      \end{center}
{\small
 Note.
 $\pi$ in mas,
 $\mu^*_\alpha=\mu_\alpha\cos\delta$ and $\mu_\delta$ in mas/yr,
 $V_r=V_r(LSR)$ in km/s, (1)~Zhang et al.~(2013).
          }
      \end{table}

 \begin{figure}[p]
 {\begin{center}
 \includegraphics[width=150mm]{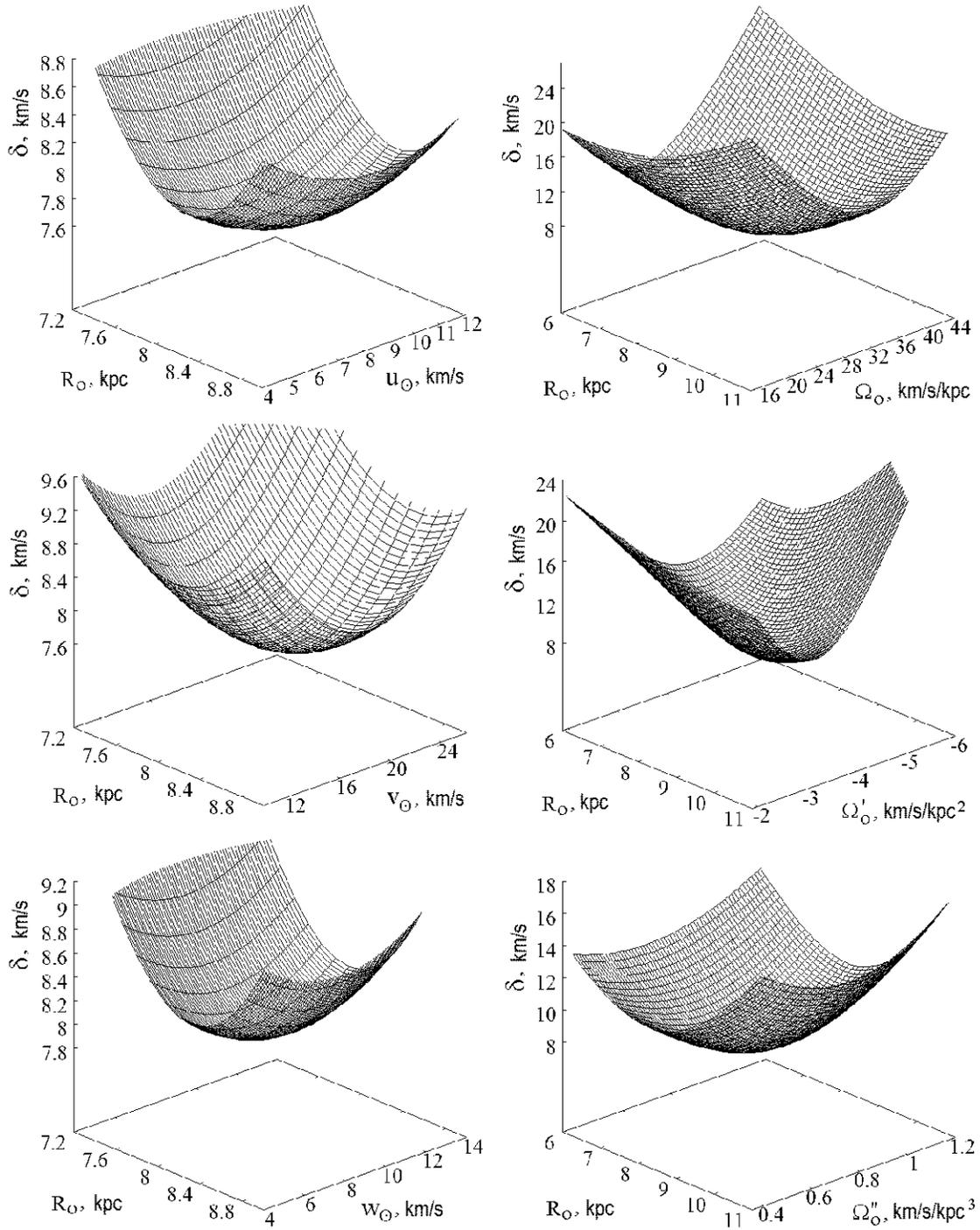}
 \caption{
Graphical representation of the two-dimensional residuals
$\delta=\sqrt{F}$ corresponding to solution (9); one of the
measurements is specified by the parameter $R_0$; one of the
parameters $u_\odot, v_\odot, w_\odot, \Omega_0, \Omega'_0,$ and
$\Omega''_0$ acts as the second measurement; the remaining
parameters are fixed at the level of the solutions obtained.
 }
 \label{delta}
 \end{center} }
 \end{figure}

\section*{RESULTS}
Several approaches to solving the optimization problem (1) with
constraints (2)–(4) are known. Since the contribution of Eq. (4)
to the general solution is negligible when using distant stars,
only Eqs. (2) and (3) may be used. In this case, however, the
velocity $w_\odot$ is determined poorly; it should be fixed. This
method was applied, for example, by Mishurov and Zenina (1999) and
Mel’nik et al. (2001).

The local expansion/contraction parameter of the stellar system K
is of great importance in analyzing nearby stars associated with
the Gould Belt (Bobylev and Bajkova 2013a), where its manifests
itself as a positive (expansion) kinematic $K$-effect for young
stars within 0.6 kpc of the Sun. A small negative $K$-effect
(contraction) manifests itself in the kinematics of various
samples of stars within up to 2 kpc of the Sun (Torra et al. 2000;
Rybka 2004; Bobylev et al. 2009). No evidence for significant
general expansion/contraction of the entire Galaxy has been
revealed, except for the region close to the Galactic center where
an expanding 3-kpc spiral arm is located (Burton 1988; Dame and
Thaddeus 2008; Sanna et al. 2009). Therefore, including the
parameter $K$ as an unknown one when analyzing masers is of
considerable interest.

Table 2 gives the kinematic parameters found by using the
three-dimensional $(V_r, V_l, V_b)$ velocity field of 73 masers.
Table 3 gives the kinematic parameters found by using the
two-dimensional $(V_r, V_l)$ velocity field of the same masers.
The solutions obtained both with unit weights and with weights (6)
are presented in the tables. The solutions obtained at a fixed
solar velocity, $w_\odot = 7$~km s$^{-1}$, are given in the last
two columns of the tables. Analysis of the parameters presented in
Tables 2 and 3 allows a number of conclusions to be reached. Since
the local expansion/contraction parameter of the stellar system
$K$ does not differ significantly from zero in all eight
solutions, the number of unknown parameters can be reduced.
Obviously, the peculiar solar velocity $w_\odot$ should be fixed
when analyzing the two-dimensional velocity field. However, this
quantity is determined with confidence in the case of a
three-dimensional analysis. This is because the sample of masers
contains a sufficient number of nearby sources. Using the system
of weights (6) slightly improves the accuracy of the parameters
being determined compared to unit weights. $R_0$ is determined
with confidence in all eight cases.

We obtained a solution using the three-dimensional maser velocity
field with seven unknowns, without the parameter $K,$ and with
unit weights:
 \begin{equation}
 \begin{array}{lll}
  u_\odot= 6.72\pm1.02~\hbox{km s$^{-1}$}, \\
  v_\odot=17.68\pm0.65~\hbox{km s$^{-1}$}, \\
  w_\odot= 7.89\pm0.35~\hbox{km s$^{-1}$}, \\
  \Omega_0= 27.81\pm0.79~\hbox{km s$^{-1}$ kpc$^{-1}$},    \\
  \Omega'_0= -3.78\pm0.19~\hbox{km s$^{-1}$ kpc$^{-2}$},\\
 \Omega''_0=  0.756\pm0.050~\hbox{km s$^{-1}$ kpc$^{-3}$},\\
   R_0= 8.35\pm0.30~\hbox{kpc}
 \label{rez-1}
 \end{array}
 \end{equation}
with the error per unit weight $\sigma_0 = 8.10$~km~s$^{-1}$.

Next, we obtained a similar solution but with weights (6). It has
the smallest error per unit weight $\sigma_0 = 7.47$ km s$^{-1}$
compared to the results presented in Table 2:
 \begin{equation}
 \begin{array}{lll}
  u_\odot= 7.81\pm0.63~\hbox{km s$^{-1}$}, \\
  v_\odot=17.47\pm0.33~\hbox{km s$^{-1}$}, \\
  w_\odot= 7.73\pm0.23~\hbox{km s$^{-1}$}, \\
  \Omega_0= 28.86\pm0.45~\hbox{km s$^{-1}$ kpc$^{-1}$},    \\
  \Omega'_0= -3.96\pm0.09~\hbox{km s$^{-1}$ kpc$^{-2}$},\\
 \Omega''_0=  0.790\pm0.027~\hbox{km s$^{-1}$ kpc$^{-3}$},\\
   R_0= 8.33\pm0.20~\hbox{kpc}.
 \label{rez-2}
 \end{array}
 \end{equation}
In this case, the linear rotation velocity at the solar distance
$R_0$ is
 $V_0=241\pm7$ km s$^{-1}$ and the Oort constants
 $A=0.5R_0\Omega_0'$ and
 $B=\Omega_0+0.5R_0\Omega_0'$ are
 $A=-16.49\pm0.60$~km s$^{-1}$ kpc$^{-1}$ and
 $B= 12.37\pm1.12$~km s$^{-1}$ kpc$^{-1}$.

As a clear illustration of the uniqueness of the solution obtained
(i.e., the existence of a global minimum of the functional F in a
wide range of sought for parameters), Fig. 1 presents the
two-dimensional dependences of the residuals $\delta=\sqrt{F}$
(see (1)) on $R_0$ and one of the parameters $u_\odot, v_\odot,
w_\odot, \Omega_0, \Omega'_0,$ and $\Omega''_0$, provided that the
remaining parameters are fixed at the level of solutions (9).

Figure 2 presents the Galactic rotation curve constructed with
parameters (9) using the value of $R_0=8.3$~kpc found; when
calculating the boundaries of the confidence region, we took into
account the uncertainty in estimating $R_0$ of 0.2 kpc.

\begin{table}[t]                                                
\caption[]{\small\baselineskip=1.0ex
 Kinematic parameters found by
 using the three-dimensional ($V_r,V_l,V_b$) velocity field of 73 masers
  }
\begin{center}
      \label{t1}
\begin{tabular}{|l|r|r|r|r|r|}\hline
 Parameters                  & $w_{r,l,b}=1$  & $w_{r,l,b}\neq1$ & $w_{r,l,b}=1$ & $w_{r,l,b}\neq1$ \\\hline

 $u_\odot,$    km s$^{-1}$   & $ 6.82\pm0.99$ & $ 7.85\pm0.55$ & $ 6.80\pm1.03$ & $ 7.81\pm0.52$ \\
 $v_\odot,$    km s$^{-1}$   & $17.51\pm0.61$ & $17.32\pm0.43$ & $17.56\pm0.62$ & $17.33\pm0.42$ \\
 $w_\odot,$    km s$^{-1}$   & $ 7.90\pm0.32$ & $ 7.68\pm0.23$ & --- & --- \\

 $\Omega_0,$     km s$^{-1}$ kpc$^{-1}$ & $28.00\pm0.85$ & $28.85\pm0.50$ & $28.02\pm0.83$ & $28.83\pm0.41$ \\
 $\Omega^{'}_0,$ km s$^{-1}$ kpc$^{-2}$ & $-3.86\pm0.26$ & $-3.91\pm0.10$ & $-3.86\pm0.27$ & $-3.90\pm0.11$ \\
$\Omega^{''}_0,$ km s$^{-1}$ kpc$^{-3}$ & $ 0.78\pm0.07$ & $ 0.77\pm0.03$ & $ 0.78\pm0.07$ & $ 0.77\pm0.03$ \\

          $K,$   km s$^{-1}$ kpc$^{-1}$ & $ 0.02\pm0.38$ & $-0.24\pm0.18$ & $-0.04\pm0.44$ & $-0.28\pm0.20$ \\
        $R_0,$   kpc                    & $ 8.25\pm0.41$ & $ 8.42\pm0.16$ & $ 8.25\pm0.43$ & $ 8.46\pm0.19$ \\
   $\sigma_0,$   km s$^{-1}$            &          8.09  &         7.74   &         8.11   &          7.75  \\\hline

\end{tabular}
\end{center}
\caption[]{\small\baselineskip=1.0ex
 Kinematic parameters found by
 using the two-dimensional ($V_r,V_l,V_b$) velocity field of 73 masers
  }
\begin{center}
      \label{t2}
\begin{tabular}{|l|r|r|r|r|r|}\hline
 Parameters                  & $w_{r,l,b}=1$  & $w_{r,l,b}\neq1$ & $w_{r,l,b}=1$ & $w_{r,l,b}\neq1$ \\\hline

 $u_\odot,$    km s$^{-1}$           & $ 4.47\pm1.27$ & $ 5.10\pm0.63$ & $ 6.53\pm1.15$ & $ 7.45\pm0.49$ \\
 $v_\odot,$    km s$^{-1}$           & $17.37\pm0.58$ & $17.37\pm0.36$ & $17.44\pm0.53$ & $17.31\pm0.41$ \\
 $w_\odot,$    km s$^{-1}$           & $36.78\pm4.86$ & $34.91\pm3.38$  & --- & --- \\

 $\Omega_0,$     km s$^{-1}$ kpc$^{-1}$ & $27.45\pm0.85$ & $27.90\pm0.42$ & $27.91\pm0.96$ & $28.71\pm0.45$ \\
 $\Omega^{'}_0,$ km s$^{-1}$ kpc$^{-2}$ & $-3.74\pm0.22$ & $-3.80\pm0.10$ & $-3.81\pm0.28$ & $-3.89\pm0.11$ \\
$\Omega^{''}_0,$ km s$^{-1}$ kpc$^{-3}$ & $ 0.75\pm0.06$ & $ 0.76\pm0.03$ & $ 0.76\pm0.07$ & $ 0.77\pm0.03$ \\

          $K,$   km s$^{-1}$ kpc$^{-1}$ & $-0.17\pm0.38$ & $-0.31\pm0.18$ & $-0.13\pm0.41$ & $-0.29\pm0.20$ \\
        $R_0,$   kpc                    & $ 8.35\pm0.38$ & $ 8.41\pm0.18$ & $ 8.37\pm0.45$ & $ 8.44\pm0.20$ \\
   $\sigma_0,$   km s$^{-1}$            &          8.07  &         7.30   &         8.45   &          7.70  \\\hline

\end{tabular}
\end{center}
\end{table}

 \begin{figure}[t]
 {\begin{center}
 \includegraphics[width=120mm]{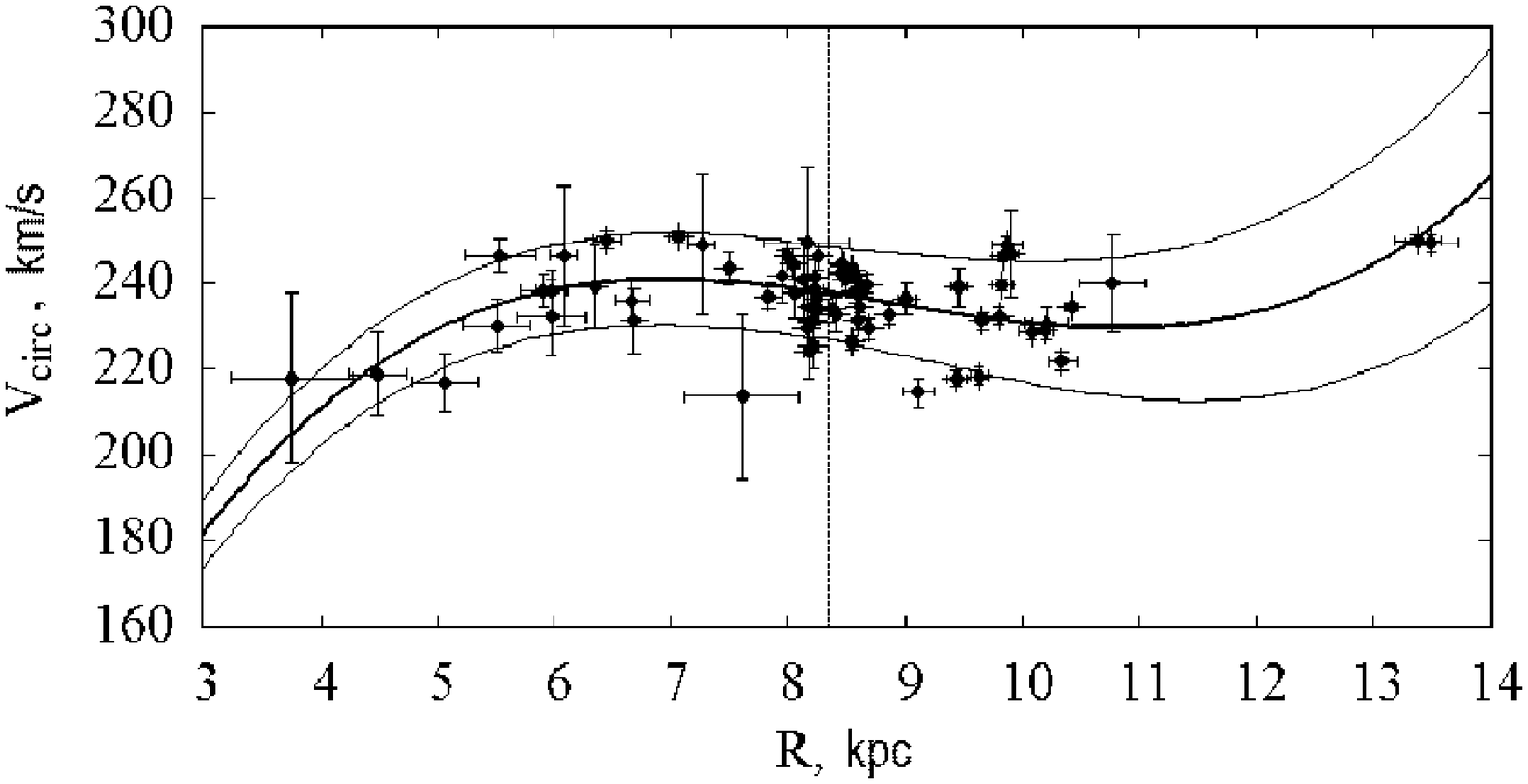}
 \caption{
Galactic rotation curve constructed with parameters (9) (thick
line); the thin lines mark the 1у confidence region; the vertical
straight line marks the Sun’s position. }
 \label{rotcurve}
 \end{center} }
 \end{figure}

 \subsection*{Allowance for the Statistical Properties of the Sample of Masers}
Although the sample of masers with measured trigonometric
parallaxes is small, we can set the objective to refine the
parameters $S_0,$ $\beta$ and $\gamma,$ needed for the calculation
of weights (6).

For this purpose, we used the well-known relations between the
dispersions $\sigma_U,\sigma_V,$ and $\sigma_W$ of the velocities
$U,V,$ and $W,$ respectively, directed along the heliocentric
rectangular coordinate axes and the dispersions
$\sigma_{V_r},\sigma_{V_l},$ and $\sigma_{V_b}$ of the velocities
$V_r,V_l,$ and $V_b,$ respectively, directed along the observed
axes (Cr\'ez\'e and Mennessier 1973):
 \begin{equation}
 \begin{array}{lll}
  \sigma^2_{V_r}=\sigma^2_U\cos^2 l\cos^2 b + \sigma^2_V\sin^2 l\cos^2 b+\\
       \qquad\sigma^2_W\sin^2 b +\varepsilon^2_{V_r},\\
  \sigma^2_{V_l}=\sigma^2_U\sin^2 l         + \sigma^2_V\cos^2 l  +(4.74r)^2\varepsilon^2_{\mu_l},\\
  \sigma^2_{V_b}=\sigma^2_U\cos^2 l\sin^2 b + \sigma^2_V\sin^2 l\sin^2 b+\\
       \qquad\sigma^2_W\cos^2 b +(4.74r)^2\varepsilon^2_{\mu_b},
 \label{Creze}
 \end{array}
 \end{equation}
where $\varepsilon_{V_r},\varepsilon_{\mu_l},\varepsilon_{\mu_b}$
are the observational errors of the corresponding components.

Note that the velocities $U,$ $V,$ and $W$ from which the
dispersions are calculated must be the residual ones, i.e., freed
from the differential Galactic rotation and the perturbations
caused by the influence of the spiral density wave. The available
sample of masers is so far insufficient for the necessity of
allowance for both effects to be shown in practice. However,
previously (Bobylev and Bajkova 2013a), using a sample of 162
nearby O--B2.5 stars as an example, we showed (see Fig. 8 in the
cited paper) that allowance for the influence of the spiral
density wave reduces tangibly the residual velocity dispersions.
In our case, to take into account the influence of the
differential Galactic rotation, we used the parameters from
solution (9). The spiral density wave parameters were taken from
Bobylev and Bajkova (2013b), where they were also determined from
masers.

To analyze the residual velocities, we produced a sample of 55
masers satisfying the conditions $\sigma_\pi/\pi<10\%$ and
$r<3.5$~kpc, i.e., we selected sufficiently distant sources with
the most reliably measured distances. After allowance for the
differential Galactic rotation and the velocities of the
perturbations caused by the influence of the spiral density wave,
we obtained the following residual velocity dispersions:
 $\sigma_U=8.2$~km s$^{-1}$,
 $\sigma_V=7.1$~km s$^{-1}$ and
 $\sigma_W=7.3$~km s$^{-1}$.
Next, after the substitution of these values into the right parts
of Eqs. (10), we found the dispersions
 ${\overline \sigma_{V_r}}$, ${\overline \sigma_{V_l}}$ and ${\overline \sigma_{V_b}}$ for
each object. The dispersions averaged over all objects are
 $\sigma_{V_r}=1.19$~km s$^{-1}$,
 $\sigma_{V_l}=1.22$~km s$^{-1}$ and
 $\sigma_{V_b}=0.99$~km s$^{-1}$,
from which we calculated the sought-for coefficients:
 $\beta=\sigma_{V_r}/\sigma_{V_l}=0.98$ and
 $\gamma=\sigma_{V_r}/\sigma_{V_b}=1.20$.
 As we see, the
derived values of $\beta$ and $\gamma$  are close to unity. The
solution of the kinematic equations (2)--(4) for 73 masers using
the derived parameters is virtually indistinguishable from
solution (9) obtained at $\beta=1$ and $\gamma=1.$

We also considered the case where a coefficient $\gamma$ differing
significantly from unity is expected to be obtained. For this
purpose, we produced a sample of 24 nearer masers satisfying the
conditions $\sigma_\pi/\pi<10\%$ and $r<1.5$~kpc. After allowance
for the differential Galactic rotation and the velocities of the
perturbations caused by the influence of the spiral density wave,
we obtained the following residual velocity dispersions:
 $\sigma_U=6.2$~km s$^{-1}$,
 $\sigma_V=4.8$~km s$^{-1}$ and
 $\sigma_W=3.6$~km s$^{-1}$,
 from which we estimated
 $S_0=\sqrt{\sigma^2_U+\sigma^2_V+\sigma^2_W}=8.6$~km s$^{-1}$.
 Using Eqs. (10) and the subsequent
averaging of the individual dispersions, we found
$\sigma_{V_r}=0.95$~km s$^{-1}$,
 $\sigma_{V_l}=0.99$~km s$^{-1}$ and
 $\sigma_{V_b}=0.49$~km s$^{-1}$, whence
 $\beta=\sigma_{V_r}/\sigma_{V_l}=0.96$ and
 $\gamma=\sigma_{V_r}/\sigma_{V_b}=1.94$. The new
solution of the kinematic equations (2)--(4) for 73 masers with
weights (6) and the refined $S_0=8.6$~км/с, $\beta=0.96,$ and
$\gamma=1.94$ is
 \begin{equation}
 \begin{array}{lll}
  u_\odot= 8.58\pm0.54~\hbox{km s$^{-1}$}, \\
  v_\odot=17.72\pm0.42~\hbox{km s$^{-1}$}, \\
  w_\odot= 7.42\pm0.23~\hbox{km s$^{-1}$}, \\
   \Omega_0= 29.05\pm0.46~\hbox{km s$^{-1}$ kpc$^{-1}$},    \\
  \Omega'_0= -3.97\pm0.09~\hbox{km s$^{-1}$ kpc$^{-2}$},\\
 \Omega''_0=  0.790\pm0.028~\hbox{km s$^{-1}$ kpc$^{-3}$},\\
   R_0= 8.34\pm0.16~\hbox{kpc}.
 \label{rez-3}
 \end{array}
 \end{equation}
In this case, the error per unit weight is $\sigma_0=7.53$~km
s$^{-1}$. It can be seen that the differences between solutions
(9) and (11) do not exceed the $1\sigma$ error level. Thus, we
obtained close solutions, despite the fact that the weights of the
velocities $V_b$ (Eq. (4)) increased considerably.

Since the coefficients $\beta\approx1$ and $\gamma\approx1$ found
above reflect the statistical properties of the original sample of
73 masers more adequately, we consider solution (9) as the basic
one. In this case, as our additional simulations showed, a low
sensitivity of the solution to changes in the coefficients $\beta$
and $\gamma$ should be noted.

\section*{DISCUSSION}
The parameters of the Galactic rotation curve we found (see
solutions (9) and (11)) are in good agreement with the results of
analyzing such young Galactic disk objects as OB associations,
$\Omega_0 =31\pm1$ km s$^{-1}$ kpc$^{-1}$ (Mel'nik et al. 2001;
Mel'nik and Dambis 2009), blue supergiants,
 $\Omega_0=29.6\pm1.6$~km s$^{-1}$ kpc$^{-1}$ and
 $\Omega'_0=-4.76\pm0.32$~km s$^{-1}$ kpc$^{-2}$ (Zabolotskikh et
al. 2002), or distant OB3 stars ($R_0=8$~kpc),
 $\Omega_0 =31.9\pm1.1$~km s$^{-1}$ kpc$^{-1}$,
 $\Omega^{'}_0 = -4.30\pm0.16$~km s$^{-1}$ kpc$^{-2}$ and
 $\Omega^{''}_0 = 1.05\pm0.35$~km s$^{-1}$ kpc$^{-3}$ (Bobylev and Bajkova 2013a).
 The value
of $V_0=241\pm7$ km s$^{-1}$ obtained at $R_0=8.3$~kpc is in good
agreement with $V_0=254\pm16$ km s$^{-1}$ at $R_0=8.4$~kpc (Reid
et al. 2009a) and $V_0=244\pm13$ km s$^{-1}$ for $R_0=8.2$~kpc
(Bovy et al. 2009) determined from a sample of 18 masers. Note
also the paper by Irrgang et al. (2013), who proposed three
Galactic potential models constructed using data on hydrogen
clouds and masers, with the velocity $V_0$ having been found to be
close to 240~km s$^{-1}$ and $R_0\approx8.3$~kpc.

Individual independent methods give an estimate of $R_0$ with an
error of 10--15\%. Note several important isolated measurements.
Based on Cepheids and RR Lyr stars belonging to the bulge
(collected by Groenewegen et al. 2008) and using improved
calibrations derived from Hipparcos data and 2MASS photometry,
Feast et al. (2008) obtained an estimate of $R_0=7.64\pm0.21$~kpc.
Having analyzed the orbits of stars moving around a massive black
hole at the Galactic center (the method of dynamical parallaxes),
Gillessen et al. (2009) obtained an estimate of
$R_0=8.33\pm0.35$~kpc. According to VLBI measurements, the radio
source Sqr A* has a proper motion relative to extragalactic
sources of $6.379\pm0.026$ mas yr$^{-1}$ (Reid and Brunthaler
2004); using this value, Sch\"onrich (2012) found
$R_0=8.27\pm0.29$~kpc and $V_0=238\pm9$ km s$^{-1}$. Two H$_2$O
maser sources, Sgr~B2N and Sgr~B2M, are in the immediate vicinity
of the Galactic center, where the radio source Sqr A* is located.
Based on their direct trigonometric VLBI measurements, Reid et al.
(2009b) obtained an estimate of $R_0=7.9^{+0.8}_{-0.7}$~kpc.

Thus, our kinematic estimate of $R_0=8.3\pm0.2$~kpc is in good
agreement with the known estimates and surpasses some of them in
accuracy.

\section*{CONCLUSIONS}
Based on published data, we produced a sample of 73 masers with
known line-of-sight velocities and highly accurate trigonometric
parallaxes and proper motions measured by VLBI. This allowed the
maser velocity field needed to solve Bottlinger's kinematic
equations to be formed. Bottlinger’s kinematic equations we
considered relate the Galactic rotation parameters ($\Omega_0$ and
its derivatives), the solar Galactocentric distance $(R_0),$ the
object group velocity components relative to the Sun ($u_\odot,
v_\odot, w_\odot$), and the system expansion/contraction parameter
(the $K$ effect). The method of minimizing the quadratic
functional that is the sum of the weighted squares of the
residuals of measurements and model velocities was used to find
the unknown parameters. Solutions were found in the cases of both
three-dimensional and two-dimensional velocity fields for various
numbers of sought-for parameters when various weighting methods
were applied. In all cases, the $K$ effect turned out to be
statistically insignificant. We established that the solution
obtained from the three-dimensional maser velocity field for seven
sought-for parameters ($u_\odot, v_\odot, w_\odot, \Omega_0,
\Omega'_0, \Omega''_0,$ and $R_0$) corresponding to the global
minimum of the functional in a wide range of their variations is
most reliable. This solution is (9). The linear rotation velocity
at the solar distance $R_0$ is $V_0=241\pm7$~km s$^{-1}$. The
solar Galactocentric distance $R_0$ is the most important and
debatable parameter. Our value is in good agreement with the most
recent estimates and even surpasses them in accuracy.

\section*{ACKNOWLEDGMENTS}
We are grateful to the referees for their useful remarks that
contributed to an improvement of the paper. This work was
supported by the ``Nonstationary Phenomena in Objects of the
Universe'' Program P--21 of the Presidium of the Russian Academy
of Sciences.

\section*{REFERENCES}
{\small

\quad~~1. A.T. Bajkova and V.V. Bobylev, Astron. Lett. 38, 549
(2012).

2. V.V. Bobylev, A.T. Bajkova, and S.V. Lebedeva, Astron. Lett.
33, 720 (2007).

3. V.V. Bobylev, A.T. Bajkova, and A.S. Stepanishchev, Astron.
Lett. 34, 515 (2008).

4. V.V. Bobylev, A.S. Stepanishchev, A.T. Bajkova, and G.A.
Gontcharov, Astron. Lett. 35, 836 (2009).

5. V.V. Bobylev and A.T. Bajkova, Mon. Not. R. Astron. Soc. 408,
1788 (2010).

6. V.V. Bobylev and A.T. Bajkova, Astron. Lett. 39, 532 (2013a).

7. V.V. Bobylev and A.T. Bajkova, Astron. Lett. 39, 809 (2013b).

8. J. Bovy, D.W. Hogg, and H.-W. Rix, Astrophys. J. 704, 1704
(2009).

9. A. Brunthaler, M.J. Reid, K.M. Menten, et al., Astron. Nachr.
332, 461 (2011).

10. W.B. Burton, Galactic and Extragalactic Radio Astronomy, Ed.
by G. Verschuur and K. Kellerman (Springer, New York, 1988).

11. D.P. Clemens, Astrophys. J. 295, 422 (1985).

12. M. Cre\'ze\' and M O. Mennessier, Astron. Astrophys. 27, 281
(1973).

13. T.M. Dame and P. Thaddeus, Astron. Astrophys. 683, 143 (2008).

14. S. Dzib, L. Loinard, L.F. Rodriguez, et al., Astrophys. J.
733, 71 (2011).

15. M.W. Feast, C.D. Laney, T.D. Kinman, et al., Mon. Not. R.
Astron. Soc. 386, 2115 (2008).

16. T. Foster and B. Cooper, ASP Conf. Ser. 438, 16 (2010).

17. C. Francis and E. Anderson, arXiv:1309.2629 (2013).

18. S. Gillessen, F. Eisenhauer, S. Trippe, et al., Astrophys. J.
692, 1075 (2009).

19. M.A.T. Groenewegen, A. Udalski, and G. Bono, Astron.
Astrophys. 481, 441 (2008).

20. T. Hirota, T. Bushimata, Y.K. Choi, et al., Publ. Astron. Soc.
Jpn. 59, 897 (2007).

21. A. Irrgang, B. Wilcox, E. Tucker, and L. Schiefelbein, Astron.
Astrophys. 549, 137 (2013).

22. M.K. Kim, T. Hirota, M. Honma, et al., Publ. Astron. Soc. Jpn.
60, 991 (2008).

23. E.S. Levine, C. Heiles, and L. Blitz, Astrophys. J. 679, 1288
(2008).

24. Z.M. Malkin, Astron. Rep. 57, 128 (2013).

25. N.M. McClure-Griffiths and J.M. Dickey, Astrophys. J. 671, 427
(2007).

26. P.J. McMillan and J.J. Binney, Mon. Not. R. Astron. Soc. 402,
934 (2010).

27. A.M. Mel’nik, A.K. Dambis, and A S. Rastorguev, Astron. Lett.
27, 521 (2001).

28. A.M. Mel’nik and A.K. Dambis, Mon. Not. R. Astron. Soc. 400,
518 (2009).

29. Yu. N. Mishurov and I.A. Zenina, Astron. Astrophys. 341, 81
(1999).

30. T. Nagayama, T. Omodaka, T. Handa, et al., Publ. Astron. Soc.
Jpn. 63, 719 (2011).

31. I.I. Nikiforov, ASP Conf. Ser. 316, 199 (2004).

32. A.S. Rastorguev, E.V. Glushkova, A.K. Dambis, and M.V.
Zabolotskikh, Astron. Lett. 25, 595 (1999).

33. M.J. Reid, Ann. Rev. Astron. Astrophys. 31, 345 (1993).

34. M.J. Reid and A. Brunthaler, Astrophys. J. 616, 872 (2004).

35. M.J. Reid, K.M. Menten, X.W. Zheng, et al., Astrophys. J. 700,
137 (2009a).

36. M. Reid, K.M. Menten, X.W. Zheng, et al., Astrophys. J. 705,
1548 (2009b).

37. S.P. Rybka, Kinem. Fiz. Nebesn. Tel 20, 133 (2004).

38. K.L.J. Rygl, A. Brunthaler, M.J. Reid, et al., Astron.
Astrophys. 511, A2 (2010).

39. A. Sanna, M.J. Reid, L. Moscadelli, et al., Astrophys. J. 706,
464 (2009).

40. R. Sch\"onrich, Mon. Not. R. Astron. Soc. 427, 274 (2012).

41. J. Torra, D. Fern\'andez, and F. Figueras, Astron. Astrophys.
359, 82 (2000).

42. R.M. Torres, L. Loinard, A.J. Mioduszewski, et al., Astrophys.
J. 698, 242 (2009).

43. Y. Xu, J.J. Li, M.J. Reid, et al., Astrophys. J. 769, 15
(2013).

44. M.V. Zabolotskikh, A.S. Rastorguev, and A.K. Dambis, Astron.
Lett. 28, 454 (2002).

45. B. Zhang, M.J. Reid, K.M. Menten, et al., Astrophys. J. 775,
79 (2013).

}
\end{document}